\documentstyle[aps,epsf,epsfig]{revtex}
\begin{document}
\author{B. Mombelli$^{1}$, O. Kahn$^{1}$, J. Leandri$^{2}$, Y. Leroyer$^{2}$, S.
Mechkov$^{2}$ and Y. Meurdesoif$^{2}$}
\address{
$^{1}$Laboratoire des Sciences Mol\'{e}culaires,\\
Institut de Chimie de la Mati\`{e}re Condens\'{e}e de Bordeaux, UPR CNRS
9048, \\
Avenue Albert Schweitzer 33608 Pessac, France\\
$^{2}$Centre de Physique Th\'{e}orique et de Mod\'{e}lisation de Bordeaux,\\
Universit\'{e} Bordeaux I, URA CNRS 1537, \\
19 rue du Solarium, 33174 Gradignan, France}
\title{{\bf Influence of the lattice geometry on the thermodynamical properties of
two-dimensional spin systems}}
\maketitle
\begin{abstract}
Various types of mixed spin two-dimensional Heisenberg networks are
investigated by means of Monte Carlo simulations. This study aims at
interpreting quantitatively the thermodynamical properties of
two-dimensional molecule-based magnets recently synthesized. The proposed
model requires that: (i) one of the two magnetic centers has a spin large
enough to be treated as a classical spin; (ii) the zero field Hamiltonian is
isotropic; (iii) the quantum spins have only classical spins as neighbours.
The quantum Hamiltonian is then replaced by a classical one with effective
ferromagnetic interactions. The temperature dependence of both the specific
heat and magnetic susceptibility are calculated. The effect of the lattice
geometry is analysed.
\end{abstract}

\section{Introduction}

A rather large number of molecule-based magnets have been synthesized and
investigated in the last few years\cite{K-book,Gat}. They correspond to
low-dimensional magnetic systems, either quasi-one-dimensional\cite
{Kahn1,Se83,DG83,VG86,Kahn2,DC89} or more recently, quasi-two-dimensional%
\cite{Kahn3,Kahn-new}. The one-dimensional compounds are well modeled for
equally-spaced magnetic chains, alternating chains involving a unique kind
of spin carrier and two kinds of interaction pathways, mixed spin chains,
ladder-type double chains, etc...

This paper is devoted to two-dimensional Heisenberg mixed spin compounds in
which one of the spins is large enough to be treated as a classical spin and
the other is normally treated as a quantum spin. This work is motivated by
the synthesis of novel two-dimensional magnetic materials. So far, two types
of two-dimensional magnetic lattices have been described. Both have a
honeycomb-like structure. The former type is obtained using oxamate as
bridging ligand. Mn$^{2+}$ ions in octahedral surroundings are located at
the corners of the hexagons, and Cu$^{2+}$ ions in elongated tetragonal
surroundings are located at the middle of the edges. A strong
antiferromagnetic interaction is propagated between the Mn$^{2+}$ and Cu$%
^{2+}$ ions through the oxamate bridge, so that the intralayer interaction
is very large as compared to the interlayer interaction. In these compounds
the layers are negatively charged, and the nature and magnitude of the
interlayer interaction is governed by the size of the countercations
situated between the layers. In the case of the NBu$^{4+}$ countercation a
long range magnetic transition was observed at 15 K, probably due to the
synergy between a very weak magnetic anisotropy and a ferromagnetic
interlayer interaction\cite{nous2}. It has been shown in a previous paper%
\cite{nous1} that our model leads to an excellent interpretation of the
magnetic properties of this material. The parameters were found as $J=$ 47.6
K, $g_{Mn}$ = 2.0 and $g_{Cu}$ = 2.2. These values are very close to those
obtained for both Cu$^{2+}$Mn$^{2+}$ pairs \cite{Kahn1} and chains \cite
{Kahn2} involving the same bridge. The value of the interaction parameter
essentially depends on the nature of the bridging network, and is not very
sensitive to the spin geometry.

The latter type of lattice is realised in a series of two-dimensional
oxalate-bridged bimetallic compounds which have just been synthesized\cite
{Kahn-new}. In that case the two kinds of magnetic ions alternate at the
corners of the hexagons. The layers which are again negatively charged are
separated by countercations. The general formula of these componds is (NBu$%
_{4}$)[M$^{II}$Ru$^{III}$(ox)$_{3}$] where M$^{II}$ is either Mn$^{II}$ ($S=$%
5/2) or Fe$^{II}$ ($S=$2) or Cu$^{II}$ ($S=$1/2) and the spin carried by the
high-field Ru$^{3+}$ ion is $S$(Ru) = 1/2.\ The interaction is ferromagnetic
for Mn and antiferromagnetic for Cu and Fe.\ In the latter case a
ferromagnetic transition occurs at $T_{c}=13$K. Let us mention that other
compounds with the same structure have been reported\cite
{TZ92,AS93,DS93,DS94}. The nature of the spin carriers, however, does not
allow to use the classical-quantum spin approach.
\begin{center}
\begin{figure}
\mbox{\epsfig{file=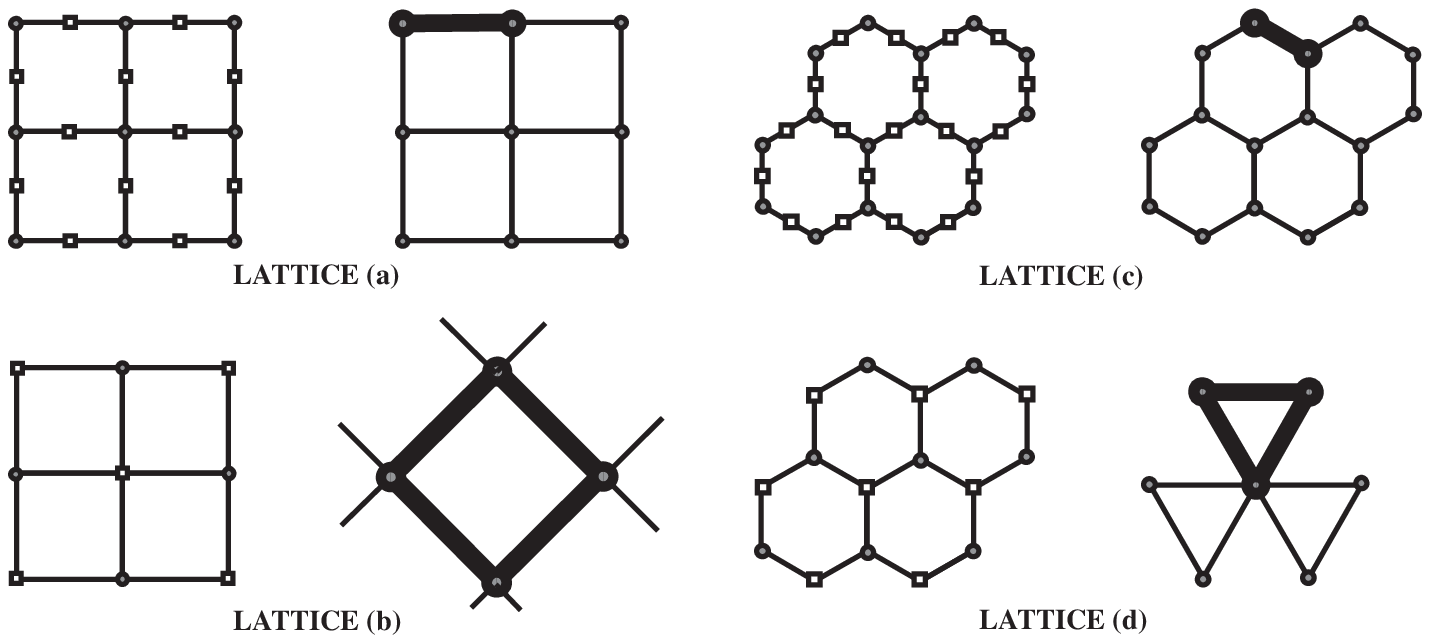}}\\
\caption{ For each cases, on the left hand side, the original lattice, where
the squares stand for quantum spins, and the circles for classical spins. On
the right hand side, the effective lattice of {\em classical} spins, with in
bold the unit cell.}
\end{figure}
\end{center}
The aim of this paper is to study the influence of the geometry of the spin
lattice on the thermodynamical properties of such systems. In this way, we
derive from the mixed quantum-classical Heisenberg model a purely classical
one with an effective interaction which depends on the lattice geometry.
This model is analysed by means of Monte Carlo (MC) simulations. We
investigate the different realisations of the hexagonal lattice described
above, but also two configurations based on the square lattice.

The paper is organized as follows: the model is developed in the first
section, then the Monte Carlo analysis is presented, and the results are
discussed in the last section.

\section{The model}

Let us write down the Heisenberg spin Hamiltonian as: 
\[
{\cal H}=J\sum\limits_{\left\langle ij\right\rangle }{\bf S}_{i}^{(Q)}\cdot 
{\bf S}_{j}^{(C)}-g_{1}\mu
_{B}H\sum\limits_{j=1}^{N_{C}}S_{j}^{z(C)}-g_{2}\mu
_{B}H\sum\limits_{i=1}^{N_{Q}}S_{i}^{z(Q)} 
\]
Here ${\bf S}_{j}^{(C)}$ is the large spin operator (5/2 for Mn, 2 for Fe)
which will be approximated by a classical vector, $S{\bf s}$ where ${\bf s}$
is a unit vector and $S=\sqrt{S^{(C)}(S^{(C)}+1)}$.The $1/2$ spin quantum
operator (for Cu or Ru) is denoted by $S_{i}^{(Q)}=\frac{1}{2}{\bf \sigma }%
_{i}$ with ${\bf \sigma }_{i}$ the Pauli matrices.\ The interaction
parameter $J$ is positive for an antiferromagnetic interaction (in the
following, we only consider this case); $H$ is a weak magnetic field applied
along the $z$ direction. $<ij>$ stands for a pair of nearest neighbour
spins, $N_{C}$ is the number of classical spins and $N_{Q}$ is the number of
quantum spins.

Two kinds of magnetic lattices are investigated, the hexagonal one, which
has been realised experimentally, and the square one. For each lattice, the
spins can be arranged in two fashions: either the classical spins are
attached at each vertex, and the quantum spins occupy the middle of the
links, or the classical and quantum spins alternate at the vertices of the
lattice. The structures are schematized in figure 1 (a) to (d).

In all cases, a quantum spin is surrounded only by classical ones.
Therefore, the partition function can be factorized with respect to the
quantum spin operators: 
\[
Z(T,H)=\int \left( \prod_{j=1}^{N_{C}}d\Omega _{j}\right) ~\text{Tr}_{\sigma
}\left\{ \prod_{i=1}^{N_{Q}}\exp \left( {\bf \sigma }_{i}\cdot [-\frac{1}{2}%
\beta JS\,%
\mathop{\textstyle \sum }%
_{j\in V(i)}{\bf s}_{j}+\frac{1}{2}\beta g_{2}\mu _{B}H\widehat{\,{\bf e}}%
_{z}]\right) \right\} ~\exp \left( \beta g_{1}\mu _{B}S\,H{\bf \cdot }%
\sum_{j=1}^{N_{C}}s_{j}^{z}\right) 
\]
In this expression, $V(i)$ is the set of labels of the classical spins
nearest neighbours of the quantum spin at site $i$. Let us call this set of
classical spins a {\em unit cell}. Depending on the lattice, the unit cell
is a link (lattices (a) and (c) of figure 1), a triangular plaquette
(lattice (d)) or a square plaquette (lattice (b)). Due to the factorized
form of $Z(T,H)$, the quantum spin dependence can be traced out to give a
fully classical partition function: 
\begin{equation}
Z(T,H)=\int \left( \prod_{j=1}^{N_{c}}d\Omega _{j}\right) ~\left\{
\prod_{\{\Gamma \}}2\cosh \parallel -\frac{1}{2}\beta JS\,%
\mathop{\textstyle \sum }%
_{j\in \Gamma }{\bf s}_{j}+\frac{1}{2}\beta g_{2}\mu _{B}H\widehat{\,{\bf e}}%
_{z}\parallel \right\} ~\exp \left( \beta g_{1}\mu _{B}S\,H{\bf \cdot }%
\sum_{j=1}^{N_{c}}s_{j}^{z}\right)  \label{ZQ}
\end{equation}
where $\{\Gamma \}$ is the set of unit cells on the lattice.

The original mixed spin system is then equivalent to a classical one with an
effective ferromagnetic interaction between the classical spins on the
plaquettes $\{\Gamma \}$ which becomes (in zero field): 
\[
{\cal H}_{\text{eff}}=-k_{B}T\sum\limits_{\{\Gamma \}}\ln \left( 2\cosh
\parallel K\,%
\mathop{\textstyle \sum }%
_{j\in \Gamma }{\bf s}_{j}\parallel \right) 
\]
In the following, we shall use the notations 
\[
K=\frac{1}{2}\frac{JS}{k_{B}T}\quad ;\quad {\bf W}(\Gamma )=%
\mathop{\textstyle \sum }%
_{j\in \Gamma }{\bf s}_{j}\quad ;\quad W(\Gamma )=\parallel {\bf W}(\Gamma
)\parallel \quad ;\quad W_{z}(\Gamma )={\bf W}(\Gamma )\cdot \widehat{\,{\bf %
e}}_{z} 
\]

The various observables, like the heat capacity $C_{V}=k_{B}\beta ^{2}\frac{%
\partial ^{2}}{\partial \beta ^{2}}\ln Z(T,0)$ and the zero field
suceptibility ${\chi =}\left. {\frac{k_{B}T}{V}\frac{\partial ^{2}}{\partial
H^{2}}\ln Z}\right| _{H=0}$ are simple generalisations of the ones defined
in ref.\cite{nous1} and can be expressed as ensemble averages with respect
to the Boltzmann weight ${e^{-\beta {\cal H}_{{\rm eff}}}/{Z(T,0)}}$. By
defining ${E=-\frac{1}{2}JS\sum_{\{\Gamma \}}W(\Gamma )~\tanh (K\,W(\Gamma ))%
}$, we find that the internal energy and the specific heat are given by 
\begin{eqnarray}
U &=&\left\langle E\right\rangle _{{\cal H}_{{\rm eff}}}  \nonumber \\
C_{V} &=&k_{B}~\beta ^{2}~\left[ \left\langle E^{2}\right\rangle _{{\cal H}_{%
{\rm eff}}}-\left\langle E\right\rangle _{{\cal H}_{{\rm eff}}}^{2}\right]
+k_{B}\left\langle \Phi \right\rangle _{{\cal H}_{{\rm eff}}}\quad \text{with%
}\quad \Phi =\sum_{\{\Gamma \}}\left[ \frac{K\,W{(\Gamma )}}{\cosh (K\,W{%
(\Gamma )})}\right] ^{2}  \label{Cv}
\end{eqnarray}
The molar magnetic susceptibility is obtained from eq.(\ref{ZQ}): 
\begin{equation}
\chi =\frac{\mu _{B}^{2}}{k_{B}T\;N_{M}}\left( g_{1}^{2}S^{2}\;\left\langle
P\right\rangle _{{\cal H}_{{\rm eff}}}+Sg_{1}g_{2}\;\left\langle
Q\right\rangle _{{\cal H}_{{\rm eff}}}+\frac{1}{4}g_{2}^{2}\;\left\langle
R\right\rangle _{{\cal H}_{{\rm eff}}}\right)  \label{chi}
\end{equation}
where $N_{M}$ is the number of molecules and $P$, $Q$, and $R$ have the
following expression\footnote{%
The numerical values of $\left\langle P\right\rangle _{{\cal H}_{{\rm eff}}}$%
, $\left\langle Q\right\rangle _{{\cal H}_{{\rm eff}}}$ and $\left\langle
R\right\rangle _{{\cal H}_{{\rm eff}}}$ as a function of the temperature for
the four lattices can be provided on request.}: 
\begin{eqnarray*}
P &=&\left( \sum\limits_{i=1}^{N_{C}}s_{i}^{z}\right) ^{2}\qquad ;\qquad
Q=\left( \sum\limits_{i=1}^{N_{C}}s_{i}^{z}\right) \times \left(
\sum\limits_{\{\Gamma \}}\bar{s}_{z}(\Gamma )\right) \\
R &=&\left( \sum\limits_{\{\Gamma \}}\bar{s}_{z}(\Gamma )\right)
^{2}-\sum\limits_{\{\Gamma \}}(\bar{s}_{z}(\Gamma
))^{2}+\sum\limits_{\{\Gamma \}}\left( (1-\rho ^{2}(\Gamma ))\frac{\tanh
(K\,W(\Gamma ))}{K\,W(\Gamma )}+\rho ^{2}(\Gamma )\right)
\end{eqnarray*}
where $\rho (\Gamma )=\frac{W_{z}(\Gamma )}{W(\Gamma )}$ and $\bar{s}%
_{z}(\Gamma )=-\;\rho (\Gamma )\,\tanh (K\,W(\Gamma ))$
\section{Monte Carlo simulations}
These various thermodynamical quantities are determined by Monte Carlo
sampling, with respect to the Boltzmann weight ${e^{-\beta {\cal H}_{{\rm eff%
}}}/{Z(T,0)}}$. The simulations were first performed using the metropolis
method \cite{MC}. This method has been proved to be very useful at high
temperature, far from the phase transition. However, it suffers from a
severe slowing down near the phase transition, and therefore becomes rather
inefficient at low temperature, since, for an isotropic two-dimensional
system, the critical temperature is $T_{C}=0\,$K\cite{MW}. A cluster
flipping method developed by Wolff \cite{Wolff} drastically reduces the
slowing down, and the Wolff algorithm was used for high values of $K$ (i.e.
low values of $T$). To overcome the finite size effects in the critical
region, we increased the size of the systems as $T$ decreased. The number of
cells is 2$^{14}$ for $K<2$, and increases up to 2$^{16}$ for $K=5$.
Periodic boundary conditions were imposed to the system, and a random spin
configuration was taken as an initial spin configuration. The number of
Metropolis steps necessary to reach the termal equilibrium was found to be
between 10$^{2}$ and 10$^{4}$ lattice sweeps depending on the temperature.
The averaging of the various observables was stopped when $\Delta \chi /\chi
<0.01$. The relative uncertainty on the energy and the specific heat was
then better than 10$^{-2}$. These calculations were performed on a Cray J916.
\begin{center}
\begin{figure}
\mbox{\epsfig{file=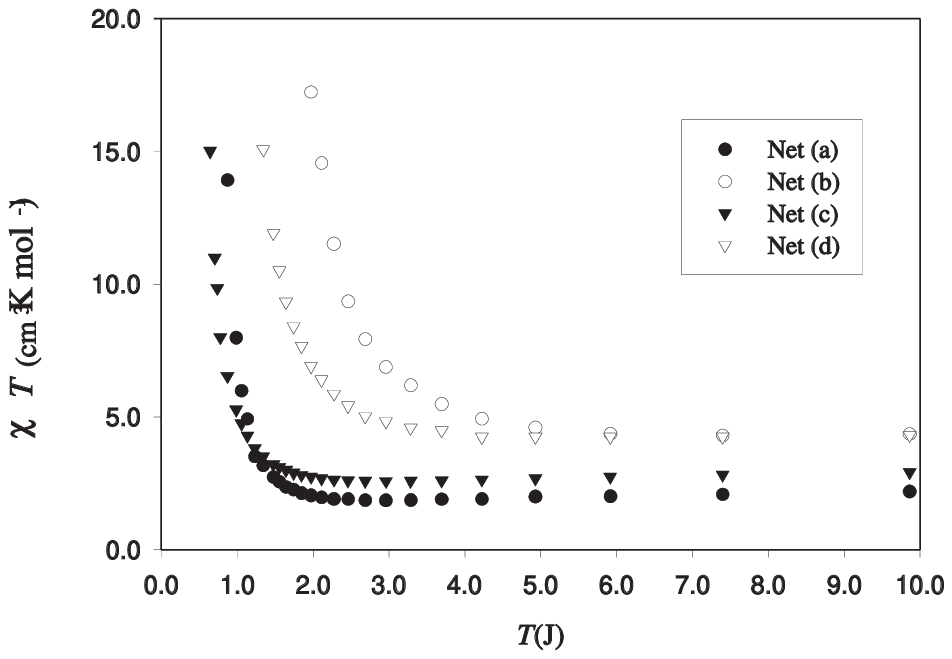}}\\
\caption{ $\chi T$ (in cm$^{3}$ K mol$^{-1}$ units) as a
function of temperature (in units of $J$). The local spins are assumed to be 
$S_{C}=S$(Mn) = 5/2 and $S_{Q}=S$(Cu) = 1/2, and the local Zeeman factors $%
g_{1}=$ $g_{2}=2$. The labels refer to the lattices of figure 1.}
\end{figure}
\end{center}
\subsection{The Magnetic Susceptibility}
Figure 2 shows the behaviour of $\chi T$ (in cm$^{3}$ K mol$^{-1}$ units) as
a function of the temperature (in units of $J$) for the four lattice
configurations.\ The four curves present the same general trend
characterised by the following features:
\begin{itemize}
\item  a constant value at high temperature corresponding to the
paramagnetic limit;
\item  a shallow minimum, characteristic of a ferrimagnetic system with
antiferromagnetic couplings.\ It is due to the local ordering appearing as
the temperature is lowered, and which causes a decrease of the local
magnetisation;
\item  a rapid increase at low temperature due to the critical divergence at 
$T=0$.
\end{itemize}
It is interesting to determine the extent to which all these results can be
described by the same universal trend, simply corrected by the lattice
effects. Actually, at high temperature, the susceptibility is given by the
Curie constant which depends in a simple way on the spin arrangement: 
\begin{equation}
\left( \chi T\right) _{T=\infty }=\frac{\mu _{B}^{2}}{3k_{B}}\left[ \frac{35%
}{4}\,n_{C}\,\,g_{1}^{2}+\frac{3}{4}\,n_{Q}\,g_{2}^{2}\,\right]
\label{Curie}
\end{equation}
where $n_{Q}$ and $n_{C}$ are the numbers of low spin (quantum) and high
spin (classical) ions per molecule, respectively (these numbers are given in
table I for each lattice).\ In equation (\ref{Curie}) we have taken $%
S^{(Q)}=1/2$, $S^{(C)}=S(Mn)=5/2$. For the curves of figure 2 we set $%
g_{1}=g_{2}=2.$
\begin{table}[h] \centering%
\begin{tabular}{cccc}
lattice & $a_{2}$ & $a_{4}$ & $a_{6}$ \\
\hline 
(a) & 4 & -$\frac{20}{3}$ & 8 \\ 
(b) & 4 & -$4$ & -$\frac{152}{9}$ \\ 
(c) & 3 & -5 & 6 \\ 
(d) & 3 & -7 & $\frac{140}{9}$%
\end{tabular}
\caption{Coefficients of the high temperature expansion for $%
c_{v}$ for the four lattices.}
\end{table}%
The temperature where the minimum occurs is measured from the Monte-Carlo
data. We present the results in table I (column 4) for each lattice. We have
computed this quantity in the mean field theory, and we found the
surprinsingly simple result : $T_{m}^{\chi }(MF)=\frac{1}{3}JS^{2}n$, where $%
n$ is the number of classical spins in a unit cell (or equivalently the
number of classical neighbours of a quantum spin).\ These values are
presented in the 5th column of table I. Although the mean field and
Monte-Carlo results are quantitatively different, one can see from table I
that the relative position of the minima with respect to the lattice is the
same for the two results. This is an indication that the parameter $n$ is
probably relevant to this quantity.

At low temperature, the critical divergence is described by the non linear
sigma model.\ The mapping between the discrete model on a specific lattice
and the universal field theory can be established as follows.\ First, take
the low temperature spin wave limit of the lattice model. 
\[
{\cal H}_{\text{eff}}(J)\longrightarrow {\cal H}_{\text{SW}}(J^{*})=-\frac{1%
}{2}J^{*}\sum_{<ij>}\theta _{ij}^{2} 
\]
Then take the long wave length limit in order to go to the continuum model 
\begin{equation}
{\cal H}_{\text{SW}}(J^{*})\longrightarrow {\cal H}_{\sigma }(\widetilde{J})=%
\frac{1}{2}\widetilde{J}\;\int d^{2}r%
\mathop{\textstyle \sum }%
_{\mu }\partial _{\mu }\,{\bf n(r)\cdot }\partial _{\mu }\,{\bf n(r)}
\label{NLSM}
\end{equation}
The first step depends on the effective interaction between the classical
spins whereas the second is related to the geometry of the classical spin
lattice. The expression of $\widetilde{J}$ is given in table I for each
lattice model. As a consequence, we expect a universal behaviour of the low
temperature regime of all the lattices, provided the temperature is
renormalised in such a way that $\widetilde{T}=T/\widetilde{J}$. We do not
observe this behaviour {\em quantitatively }on our Monte-Carlo data, since
our lowest temperatures do not lie within the universal critical regime%
\footnote{%
we noticed in ref\cite{nous1} that this behaviour is observed for lattice
(c) for $K\gtrsim 2.5$ which corresponds to $T\lesssim 0.59J$}.\ However,
the hierarchy of the $\widetilde{J}$ values gives qualitatively well the
relative positions of the critical increase of $\chi T$ for each lattice.
\begin{center}
\begin{figure}
\mbox{\epsfig{file=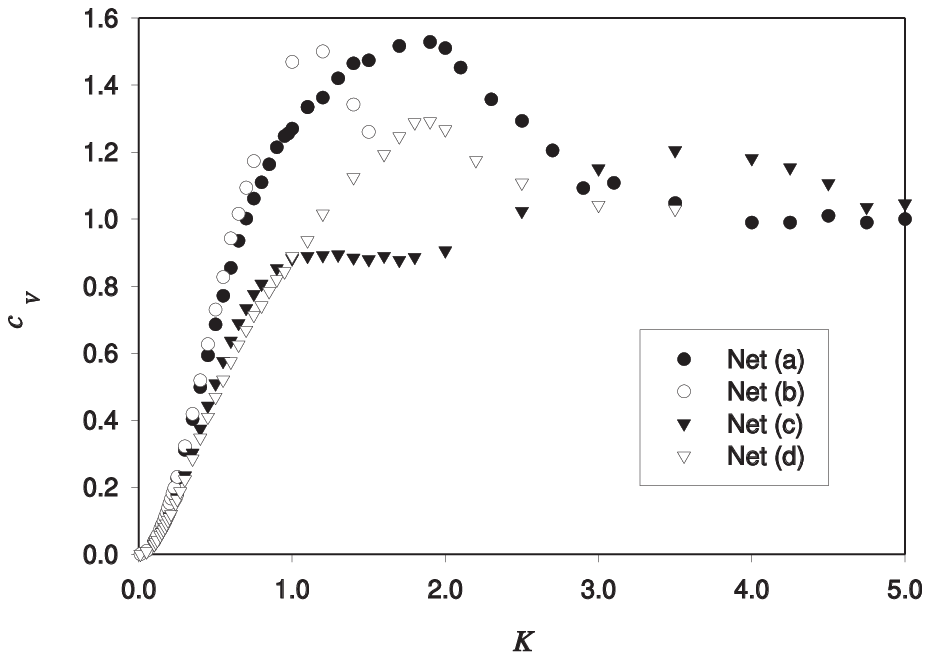}}\\
\caption{ The specific heat $c_{v}=\frac{C_{v}}{N_{c}k_{B}}$
versus $K=\frac{JS}{2k_{B}T}$ for the four lattices.}
\end{figure}
\end{center}
\subsection{The Specific Heat}
The specific heat is plotted as a function of $K$ in figure 3 for the four
types of lattices. In all cases, it presents a well pronounced maximum.
Beside this common trend, the dependence on the lattice geometry of the
details of these curves can be understood from simple arguments.\ The $T=0$
(large $K$) limiting value can be obtained from the magnon contribution to
the hamiltonian of eq.(\ref{NLSM}) which gives an energy {\em per classical
spin} 
\[
E\approx E_{0}+k_{B}T 
\]
Therefore, by normalising the heat capacity to the number of classical
spins, we get for $T=0$, $c_{V}=C_{V}/N_{C}k_{B}=1$ as can be seen on figure
3 from the result of the simulation. For large $T$ (small $K$), the
behaviour of the specific heat can be inferred from the high temperature
expansion. The first few terms of this expansion can easily be derived and
we get
\[
c_{V}=a_{2}K^{2}+a_{4}K^{4}+a_{6}K^{6}+\cdots 
\]
where $a_{2}$ turns out to be the number of unit cells connected to a single
(classical) site.\ This number together with the coefficient $a_{4}$ and $%
a_{6}$ are given in table II for each lattice The result, presented in figure
4 in comparison with the Monte-Carlo data, shows that up to $K\simeq 0.4$,
the system is driven by its high temperature behaviour.
The behaviour at intermediate and low temperature results on the
superposition of two contributions, \begin{center}
\begin{figure}[t]
\mbox{\epsfig{file=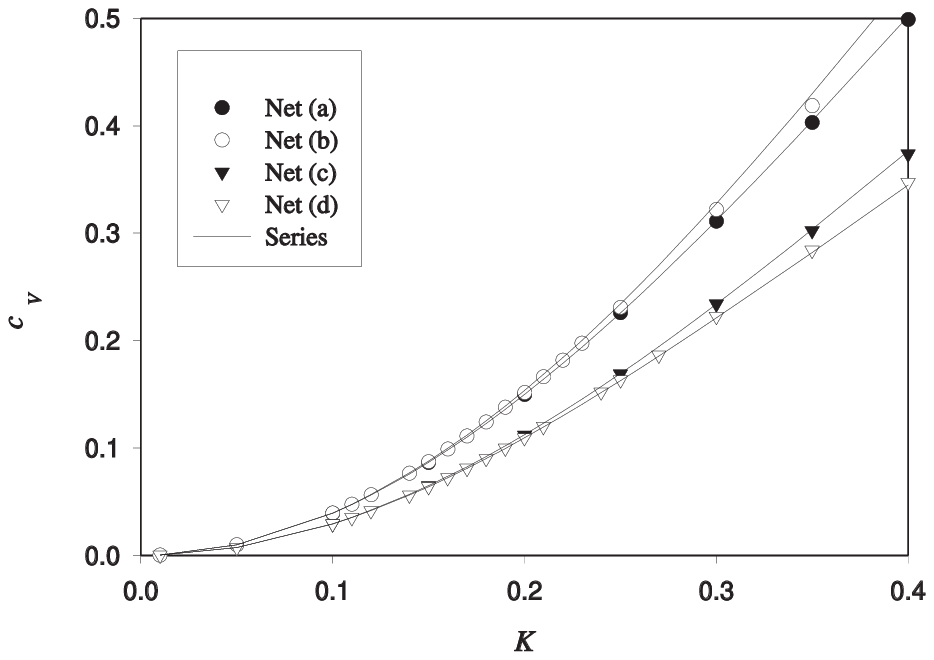}}\\
\caption{ Same as figure 3 but in the high temperature regime.
The data points correspond to the Monte-Carlo results and the lines to the
low order high temperature expansion.}
\end{figure}
\end{center}
similarly to the case of lattice (c)
already analysed in ref.(\cite{nous1}) and where this effect is clearly
visible:
\begin{itemize}
\item  a first bump for $K\approx 1$ resulting from the local ordering of
the quantum spin with respect to their randomly distributed classical
neighbours.\ By neglecting the correlations between the classical spins, we
can estimate this contribution 
\begin{equation}
C_{V}^{Q}\simeq N_{Q}k_{B}K^{2}\int 
\mathop{\textstyle \prod }%
_{i\in \Gamma }\frac{d\Omega _{i}}{4\pi }\frac{W^{2}{(\Gamma )}}{\cosh
^{2}(K\,W{(\Gamma )})}  \label{CvQ}
\end{equation}
\item  a second bump at higher value of $K$, connected to the critical
behaviour of the system and which is expected to be described by the low
temperature limit of the model (eq.(\ref{NLSM})).
\end{itemize}
\begin{table}[t] \centering%
\begin{tabular}{cccc}
lattice & $a_{2}$ & $a_{4}$ & $a_{6}$ \\
\hline 
(a) & 4 & -$\frac{20}{3}$ & 8 \\ 
(b) & 4 & -$4$ & -$\frac{152}{9}$ \\ 
(c) & 3 & -5 & 6 \\ 
(d) & 3 & -7 & $\frac{140}{9}$%
\end{tabular}
\caption{Coefficients of the high temperature expansion for $%
c_{v}$ for the four lattices.}
\end{table}%
\begin{center}
\begin{figure}[b]
\mbox{\epsfig{file=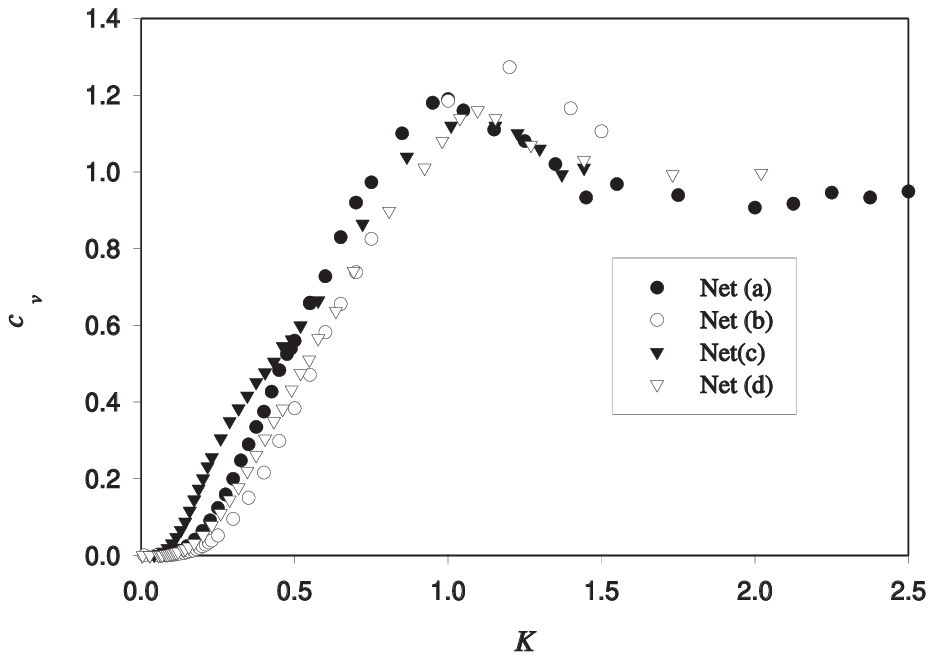}}\\
\caption{ The specific heat substracted from the quantum spin
contribution as a function of the lattice renormalised coupling $\widetilde{K%
}=\frac{\widetilde{J}}{J}K$, (see table I). }
\end{figure}
\end{center}
By substracting the specific quantum contribution given by eq.(\ref{CvQ})
and renormalising the temperature dependence of the residual specific heat
by the factor defined in the preceding section, we should obtain a universal
curve. This is what we can observe in figure 5, where the dependence of the
lattice has been washed out by comparison with figure 3.

\section{Conclusion}

In this paper we have determined the effect of the topology of the spin
lattice on the thermodynamical properties of two-dimensional systems with
alternating quantum-classical spins, modeling a wide family of magnetic
molecular compounds. Since the quantum spin dependence can be traced out, we
can used the very powerful {\em classical} Monte-Carlo techniques to analyse
these systems.

We found that the temperature dependence of both the magnetic susceptibility
and the specific heat can be described on the ground of general behaviours
on which the lattice influence is explicited. This analysis allows one to
predict in principle what would be the specific heat and the magnetic
susceptibility for other compounds with different spin geometry.
Alternatively, it can be used to determine the interaction parameters
together with the Zeeman factors out of experimental data\cite{Kahn-new}.

In most cases\cite{Kahn1}, one observes experimentally at low temperature
(typically 10 to 15K) a transition towards a ferromagnetic ordered phase
which cannot be accomodated within our isotropic interaction \cite{MW}.
Therefore, the present analysis is only valid in the paramagnetic phase. At
low temperature, spin anisotropy\cite{nous2} and/or spatial anisotropy must
be taken into account to explain this ferromagnetic phase transition.

\end{document}